\documentclass[nofootinbib,twocolumn,prd,groupedaddress]{revtex4}

\usepackage{slashed}
\usepackage{amsmath}
\usepackage{amssymb}
\usepackage{color}

\newcommand{\cmark}{\hbox{$\mathfrak{h}$}}%
\newcommand{\fmark}{\hbox{$\mathfrak{h}_F$}}%
\newcommand{\hol}{\mathfrak{h}}%
\newcommand{\nhol}{\mathfrak{n}}%

\newcommand{\weak}{\slashed{\hol}_w}
\newcommand{\strong}{\slashed{\hol}_s}

\newcommand{\Lag}{\mathcal{L}}

\def\nn{\nonumber \\ }
\def\rd{{\rm d}}

\begin{document}


\title{Holomorphy without Supersymmetry in the Standard Model Effective Field Theory}

\author{Rodrigo Alonso}
\author{Elizabeth E.~Jenkins}
\author{Aneesh V.~Manohar}

\affiliation{\vspace{1mm}
Department of Physics, University of California at San Diego, La Jolla, CA 92093, USA}


\begin{abstract}
The anomalous dimensions of dimension-six operators in the Standard Model Effective Field Theory (SMEFT) respect holomorphy to a large extent. The holomorphy conditions are reminiscent of supersymmetry, even though the SMEFT is not a supersymmetric theory. 

\end{abstract}

\maketitle

\section{Introduction}\label{sec:intro}

The Standard Model (SM) is the most general renormalizable $SU(3) \times SU(2) \times U(1)$ gauge theory built out of $n_g=3$ families of fermions and a single Higgs doublet $H$, and it has been experimentally tested in all its fundamental aspects.  In view of the absence of extra  particles at the electroweak scale, new physics effects can be included naturally by adding higher dimensional operators built with SM fields~\cite{Buchmuller:1985jz,Grzadkowski:2010es}.  This generalization of the SM defines the SMEFT built out of SM fields, which consists of the SM Lagrangian and arbitrary higher dimension operators suppressed by the scale $\Lambda$ of new physics.  Electroweak symmetry is broken spontaneously by the usual Higgs mechanism.  Any model of new physics maps to the SMEFT with specific coefficients for energies $E < \Lambda$, as long as there are no new particles present at the electroweak scale.  For energies $E < \Lambda$, the dominant new physics operators are mass dimension $d=6$.\footnote{The single $d=5$ lepton-violating operator in the SMEFT~\cite{Weinberg:1979sa} leads upon spontaneous symmetry breakdown to light Majorana masses for neutrinos which couple to the $W^\pm$ and $Z$ gauge bosons.  The extreme lightness of these neutrinos required by neutrino oscillation data implies that the energy scale of the $d=5$ operator $\Lambda_5 \gg \Lambda$.  Approximate lepton number symmetry suffices to maintain this hierarchy of new physics scales.  A similar hierarchy applies for $d=6$ operators which violate baryon number.}

In a series of papers~\cite{Grojean:2013kd,Jenkins:2013zja,Jenkins:2013wua,Alonso:2013hga,Alonso:2014zka}, we have computed the one-loop anomalous dimensions of the dimension-six operators, as well as their contributions to the anomalous dimensions of the SM $d \le 4$ parameters. The renormalization group equations (RGE) preserve gauge and flavor symmetries. Surprisingly, the one-loop RGE also preserve a holomorphic structure reminiscent of supersymmetry, even though the SMEFT is not supersymmetric. In Ref.~\cite{Alonso:2013hga}, we pointed out that the anomalous dimensions of the magnetic dipole operators preserved holomorphy.  In this paper, we summarize the non-trivial anomalous dimension conditions that are satisfied which preserve holomorphy.  We have been unable to come up with a general explanation for this holomorphic structure.   However, the large number of holomorphy relations which are satisfied suggests that this structure is not purely accidental.

Our calculations are done using the non-redundant operator basis of Ref.~\cite{Grzadkowski:2010es}, and using the equations of motion (i.e.\ field redefinitions) to reduce the operators to this standard basis. The calculation can be thought of as a computation of $S$-matrix elements, since we are computing on-shell amplitudes. The holomorphic structure only appears after this is done, with non-holomorphic direct contributions being cancelled by non-holomorphic indirect contributions from equation of motion terms. 

There have also been recent efforts to understand the form of the anomalous dimension matrix based on a tree/loop operator classification scheme~\cite{Elias-Miro:2013mua,Elias-Miro:2013gya}.

\section{Holomorphy}\label{sec:hol}

The 59 dimension-six operators can be divided into different classes depending on field content. Let $X$ denote a field-strength tensor, $\psi$ a fermion field which can be either left-handed ($L$) or right-handed ($R$), and $D$ a covariant derivative. Then the operator classes are denoted by $X^3$, $H^6$, $H^4 D^2$, $X^2H^2$, $\psi^2 H^3$, $\psi^2 H X$, $\psi^2 H^2 D$, and $\psi^4$ operators, using the notation of Refs.~\cite{Grzadkowski:2010es,Jenkins:2013zja,Jenkins:2013wua,Alonso:2013hga}.  It is convenient to separate the $\psi^4$ operators into three subclasses: $(\overline LR)(\overline LR)$,  $(\overline LR)(\overline RL)$, and current-current operators $JJ$, which consist of $(\overline LL)(\overline LL)$, $(\overline RR)(\overline RR)$, and $(\overline LL)(\overline RR)$.

To specify what we mean by holomorphic operators, we introduce the ``complex'' field strengths
\begin{align}
X^{\pm}_{\mu \nu} &=\frac12\left(  X_{\mu \nu} \mp  i \widetilde X_{\mu \nu}\right), &
\widetilde X^{\pm}_{\mu \nu} &=  \pm i X^{\pm}_{\mu \nu} \,,
\end{align}
where $\widetilde X_{\mu \nu} =  \epsilon_{\mu \nu \alpha \beta} X^{\alpha \beta}/2$, and $\epsilon_{0123}=+1$. The self-duality condition in Minkowski space is complex because $\widetilde{\widetilde X}_{\mu\nu}=-X_{\mu\nu}$.

\emph{The holomorphic part of the Lagrangian, $\mathcal{L}_\hol$, is the Lagrangian constructed from the fields $X^+$, $R$, $\overline L$, but none of their hermitian conjugates.} The Lagrangian contains also the hermitian conjugate of the holomorphic piece, $\mathcal{L}_{\bar \hol}$, which is built from the fields $X^-$, $\overline R$ and $L$. We refer to this part of the Lagrangian as anti-holomorphic. The remaining terms in the Lagrangian are deemed non-holomorphic.  

A few comments:
\begin{enumerate}

\item Under the Lorentz group $SU(2)_R \times SU(2)_L$, the fields in ${\cal L}_\hol$ transform under $SU(2)_R$, $\{ X^+$, $R$, $\overline L \} \sim \{(1,0), (\frac12,0), (\frac12,0)\}$, while the fields in ${\cal L}_{\overline \hol}$ transform under $SU(2)_L$.

\item  Holomorphy is not imposed on the Higgs doublet; in this regard the definition differs from that of supersymmetry~\cite{Intriligator:1995au} (we will come back to this point later).

\item A spinor-helicity formalism~\cite{Elvang:2013cua} study shows that holomorphic operators induce amplitudes with all particles having the same sign helicity (``$-$'' for all particles outgoing).\footnote{We thank Andrew Cohen for suggesting using the spinor-helicity method.}

\end{enumerate}

We now discuss explicitly the operators which fall into the holomorphic and anti-holomorphic categories. The $X^3$ operators are defined by
\begin{align}
Q_X &= f^{ABC} X_\mu^{A\nu} X_\nu^{B\rho} X_\rho^{C\mu},\\ \nonumber
Q_{\widetilde X}   &= f^{ABC} \widetilde X_\mu^{A\nu} X_\nu^{B\rho} X_\rho^{C\mu},
\label{1}
\end{align}
where $f^{ABC}$ are the group structure constants.  The holomorphic and anti-holomorphic combinations of these operators are
\begin{align}
Q_{X,\pm} &\equiv \frac12 \left(Q_X \mp i Q_{\widetilde X} \right) 
= f^{ABC} X^\pm_\mu{}^{A\nu} X^\pm_\nu{}^{B\rho} X^\pm_\rho{}^{C\mu} 
\end{align}
with $Q_{X,+}$ holomorphic and $Q_{X,-}$ anti-holomorphic. The contribution of the $X^3$ operators to the Lagrangian is
\begin{align}
\mathcal{L}\supset C_X Q_X + C_{\widetilde X}Q_{\widetilde X}  = C_{X,+} Q_{X,+} +   C_{X,-}Q_{X,-}
\end{align}
with complex coefficients $C_{X,\pm}\equiv\left(C_X \pm iC_{\widetilde X} \right)$.  Similarly, the $X^2H^2$ Higgs-gauge operators can be divided into holomorphic, $(X^+)^2 H^2$, and anti-holomorphic, $( X^-)^2 H^2$, operators with complex coefficients $C_{HX,\pm}\equiv\left(C_{HX} \pm iC_{H\widetilde X} \right)$.

The $\sigma_{\mu \nu}$ matrices satisfy the self-duality relation
\begin{align}
\frac{i}{2} \epsilon^{\alpha \beta \mu \nu} \sigma_{\mu \nu} P_R &= -\sigma^{\alpha \beta} P_R
\end{align}
so that the magnetic moment operators 
\begin{align}
\overline L\, \sigma^{\mu \nu} \,R X_{\mu \nu} H &=  \overline L\, \sigma^{\mu \nu} R \,X^{+}_{\mu \nu} H\,
\label{5}
\end{align}
are holomorphic, depending only on $X^+$.  Their hermitian conjugates $\overline R\, \sigma^{\mu \nu} L \, X^{-}_{\mu \nu} H^\dagger$ are anti-holomorphic.  Finally, the $(\overline LR)(\overline LR)$ operators are holomorphic. 

The $\psi^2 H^3$ operators have the form of the SM Yukawa couplings ($\widetilde H_i = \epsilon_{ij} H^{\dagger j}$)
\begin{align}
\mathcal{L} _{\rm Y} &= -  \overline q^j\, Y_d^\dagger\, d\, H_j -  \overline q^j\, Y_u^\dagger\, u\, \widetilde H_{j} - \overline l^j\, Y_e^\dagger\,  e\, H_{j}  + \hbox{h.c.}
\label{yuk}
\end{align}
multiplied by an additional factor of $H^\dagger H$. They are a priori holomorphic; however, they behave as non-holomorphic operators, and we leave them out of the holomorphic class. It is worth pointing out that these operators can be rewritten through equations of motion in terms of non-holomorphic operators, which might be the reason for their non-holomorphic behavior. 

The rest of the $d=6$ operators, $JJ$,  $H^6$,  $H^4D^2$, $\psi^2 H^2 D$ and $(\overline LR)(\overline RL)$, are also non-holomorphic.  Indeed all of them, except $(\overline LR)(\overline RL)$ and the $\psi^2 H^2 D$ operator $Q_{Hud}$, are self-conjugate, which is incompatible with any definition of holomorphy.

In summary, the Lagrangian reads:
\begin{align}
\Lag^{d=6}&=\Lag_\hol+\Lag_{\bar\hol}+\Lag_\nhol=C_\hol Q_\hol+C_{\bar\hol}Q_{\bar\hol}+C_\nhol Q_{\nhol},\\ \nonumber
Q_\hol&\subset\left\{X^3,\, X^2 H^2,\, \psi^2 X H,\, (\overline LR)(\overline LR)\right\}\\ \nonumber
Q_\nhol&\subset\left\{H^6,\,  H^4D^2,\, \psi^2 H^3,\,\psi^2 H^2 D,\,(\overline LR)(\overline RL),\, JJ \right\}
\end{align}
where $\hol$, $\bar\hol$ and $\nhol$ refer to holomorphic, anti-holomorphic and non-holomorphic operators (and their coefficients) respectively and $C_{\bar\hol}=C_\hol^*$. Note that, {\it at tree level, $\Lag_\hol$ is also holomorphic in the coefficients $C_\hol$}. 

This definition of holomorphy can be extended to the SM Lagrangian. The gauge kinetic terms can be written as a sum of holomorphic and anti-holomorphic pieces in the presence of a $\theta$-term. While the Yukawa couplings seem to be holomorphic, just like the class $\psi^2H^3$ operators, they can be rewritten in terms of non-holomorphic operators using equations of motion, so they are considered to be non-holomorphic operators.  The rest of the SM Lagrangian is self-conjugate, and non-holomorphic.

\section{RGE}
For the holomorphic part of the Lagrangian to remain so under renormalization group evolution, holomorphic operators must not mix with their hermitian conjugates or receive contributions from non-holomorphic operators. This condition ensures that $\Lag_\hol$ stays a holomorphic function of $C_\hol$ at the quantum level:
\begin{align}
C_\hol(\mu)=C_\hol\left(\left\{ C_\hol(\mu_0)\right\}, \mu_0/\mu \right),
\end{align}
where $\mu$ is the renormalization scale. We refer to this condition as the weak version of holomorphy.

This condition translates straightforwardly into the anomalous dimension matrix (note that $\gamma_{ij}$ is defined as the matrix for the coefficients $C$ rather than the operators) :
\begin{align}
\dot C_i \equiv 16 \pi^2 \mu \frac{\rd}{\rd \mu} C_i &= \sum_{j=\hol,\overline\hol,\nhol} \gamma_{ij} C_j, & i&=\hol,\overline\hol,\nhol\label{eq:rgehol}
\end{align}
where $\gamma_{ij}$ is a (non-holomorphic) function of the SM parameters and the RGE of $C_{\overline\hol}$ are the complex-conjugates of those for $C_\hol$. Weak holomorphy requires that 
$\gamma_{\hol\overline\hol}=0$ and $\gamma_{\hol \nhol}=0$, whereas it sets no constraint on $\gamma_{\nhol \hol}$.

The one-loop anomalous dimension matrix is summarized in Table~\ref{tab:1}.  It is written as the $2 \times 2$ block matrix 
\begin{align}\label{adim}
\left( \begin{array}{cc} \gamma_{\hol \hol} & \gamma_{\hol \nhol} \\ \gamma_{\nhol \hol} & \gamma_{\nhol \nhol} \end{array} \right),
\end{align}
which encodes the same information  as Eq.~(\ref{eq:rgehol}).

The one-loop anomalous dimension of the SMEFT $d=6$ operators has a number of vanishing entries; some of them are constrained to vanish by the naive dimensional analysis (NDA) perturbative order formula~\cite{Manohar:1983md,Jenkins:2013sda},  e.g.\ the $X^3-\psi^2 XH$ anomalous dimension must vanish at one loop. These cases are marked with 0 in Table~\ref{tab:1}. NDA also gives the order in coupling constants of the various entries. This information can be combined with flavor symmetry to further constrain the mixing. For example, the $(\overline LR)(\overline LR)- \left[(\overline LR)(\overline LR)\right]^\dagger$ entry of the anomalous dimension matrix is at most second order in the Yukawa couplings $Y$ by NDA;  it must vanish at one-loop order, since  flavor symmetry requires four factors of $Y$ in the mixing term. Cases where holomorphy is satisfied using a combination of NDA and flavor symmetry are marked with \fmark, signaling $\gamma_{i\bar \hol}=0$. Sometimes there do not exist any one-loop diagrams contributing to an entry either directly or indirectly through equations of motion; these entries are marked as $\nexists$.

The above considerations summarize the information one has on the anomalous dimension matrix before doing the actual computation, and they follow from general principles that apply to an arbitrary EFT. In view of Table~\ref{tab:1}, they fall far short of accounting for the structure of the anomalous dimension matrix. In the notation of Table~\ref{tab:1}, holomorphy is preserved if the $\gamma_{\hol\hol}$ block is holomorphic, i.e.\ has a \cmark, \fmark \, or vanishes in every one of its entries, and the $\gamma_{\hol \nhol}$ blocks vanishes.
The \cmark\ indicates that the entry depends only on $C_\hol$, and not on its conjugate $C_\hol^*=C_{\overline\hol}$; that is $\gamma_{i\bar \hol}=0$. The remaining symbol in Table~\ref{tab:1}, $\to 0$, signals a vanishing entry which is not expected to cancel by any of the above considerations, but which vanishes by explicit calculation.

\begin{table*}
\renewcommand{\arraycolsep}{0.15cm}
\renewcommand{\arraystretch}{2}
\begin{align*}
\begin{array}{c|cccc|cccccc}
& (X^{+})^3 & (X^{+})^2  H^2 & \psi^2 X^{+}  H  & (\overline L R) (\overline L R)  & (\overline L R) (\overline R L)  & JJ & \psi^2 H^3 & H^6 & H^4 D^2   & \psi^2 H^2 D \\
\hline
 (X^{+})^3  &  \cmark  &  \to0 & 0 &  0   & 0 & 0 & 0 &  0 & 0 & 0 \\
 (X^{+})^2 H^2  &  \cmark  &  \cmark  &  \cmark  &  0  &  0 &  \nexists & 0 & 0 & \to0 &  \to0 \\
\psi^2 X^{+}  H  &  \cmark   &  \cmark &  \cmark &  \fmark  &   \to 0 &  \to 0 & \to 0 & 0 &   \nexists &  \to0 \\
 (\overline L R) (\overline L R)  &  \to0 & \nexists &   \fmark  &  \fmark  
 & \weak: Y_u^\dagger  Y_{e,d}^\dagger & \weak: Y_u^\dagger  Y_{e,d}^\dagger   &  \nexists  &   \nexists &  \nexists  & \to0   \\
 \hline
 (\overline L R) (\overline R L)  &  \to 0 &  \nexists & \to 0 & \strong: Y_u Y_d, Y_u^\dagger Y_e^\dagger   & \fmark  & * &
 \nexists  &   \nexists &  \nexists  &  \to 0    \\
 JJ &  \to 0  & \nexists &  \to 0 & \strong: Y_u Y_{e,d}    & * & * &  \nexists  & \nexists &  \nexists   & *   \\
\psi^2 H^3 &  \to 0 & \cmark &  {\cmark} &  {\cmark} & * & *  &   \slashed{\hol}:Y_u^\dagger  Y_{e,d}^\dagger & 
 \nexists & *  & *  \\
H^6 &  \to 0 & \strong  &  \nexists  &  \nexists & \nexists & \nexists & * & * & *    & * 
 \\
H^4 D^2 & \to 0 &  \to 0 & \to 0  &  \nexists & \nexists & \nexists &  \to 0 &  \nexists & *  & *  \\
\psi^2 H^2 D & \to 0 & \to 0 &  \to 0 &  \to 0  &  \to 0 &  *  &  \to 0 & \nexists & *  &  *  \\
\end{array}
\end{align*}
\caption{\label{tab:1} Form of the one-loop anomalous dimension matrix as defined in Eqs.~(\ref{eq:rgehol},\ref{adim}) for $d=6$ operators. $Y$ is a Yukawa coupling.
The first 4 rows and columns involve holomorphic operators, and the rest involve non-holomorphic operators. 
The RGE for the rows can depend on the $C$ of each column, or their conjugates.
Entries which must vanish by NDA are denoted by $0$,  those for which there is no one-loop diagram (after taking equations of motion into account) are denoted by $\nexists$, and those which vanish by explicit computation are denoted by $\to 0$. Entries with $\cmark$ are non-zero, and satisfy holomorphy, i.e.\ they depend on $C$ but not $C^*$. Entries with  \fmark\ satisfy holomorphy because anti-holomorphic contributions are forbidden by NDA and flavor symmetry. Entries with a $*$ are non-zero. Entries with $\weak,\strong$ violate weak and strong holomorphy, respectively. The notation $\weak: Y_u^\dagger  Y_{e,d}^\dagger $, etc.,\ means that the holomorphy violation is proportional to the product $Y_u^\dagger  Y_{e,d}^\dagger$ of Yukawa couplings. The $\psi^2 H^3-\psi^2 H^3$ entry has holomorphic terms, as well as non-holomorphic terms proportional to the product $Y_u^\dagger  Y_{e,d}^\dagger$ of Yukawa couplings.}
\end{table*}

There are 16 entries in the $\gamma_{\hol \hol}$ block, all of them satisfying holomorphy.  Three entries vanish by NDA, and one entry has no one-loop graph. There are 12 remaining entries. Two respect holomorphy because they vanish by explicit computation (denoted by $\to0$), and 10 are non-vanishing but satisfy holomorphy. For example, the running of the magnetic moment operators $\dot C_{\psi^2 X H}$ depends on $ C_{\psi^2 XH}$ but not on $ C_{\psi^2 XH}^*$, as was noted in Ref.~\cite{Alonso:2013hga}, and has the entry $\cmark$. The $ C_{\psi^2 XH}^*$ term cancels between direct contributions to $\dot C_{\psi^2 X H}$ and equation of motion terms.   For holomorphy to hold for the $\dot C_{X^3} \propto C_{X^3}$ term, the $Q_X-Q_X$ anomalous dimension must equal the $Q_{\widetilde X}-Q_{\widetilde X}$ anomalous dimension, etc. 

The number of conditions that are satisfied is actually much larger than the number of entries in Table~\ref{tab:1}. Each operator class has several operators, so the entries in the table are really submatrices. In addition, many entries have several flavor invariants and/or different factors of the gauge couplings, all of which must satisfy holomorphy.\footnote{For $n_g=3$ generations, there are a total of 2499 independent real coefficients in the dimension six Lagrangian, which fall into 151 independent flavor representations. For $n_g >3$, there are 156 independent flavor representations.} 

The operators $\beta(g_X) X^2/g_X$ and $g_X^2 X \widetilde X$ are not renormalized, which implies that $X^2$ and $X \widetilde X$ have different anomalous dimensions beyond one loop. This would lead to a violation of holomorphy (e.g.\ in the $X^2 H^2$ operators, see the discussion in Ref.~\cite{Grojean:2013kd}) at two loop order. In supersymmetry, one can define a holomorphic coupling so the $\beta$ function only has a one-loop contribution~\cite{Intriligator:1995au}; this choice is necessary to preserve holomorphy beyond one loop in the $X^2 H^2$ sector.

The $\gamma_{\hol \nhol}$ block of the matrix has 24 entries, 22 of which vanish, and 2 (denoted $\weak$) which do not, violating weak holomorphy.  The two sole non-zero entries in violation of holomorphy arise from $\dot C_{(\overline LR)(\overline LR)} \propto C_{(\overline LR)(\overline RL)},C_{JJ}$.  An interesting feature of this contribution is that it is induced by a loop diagram with a virtual Higgs doublet exchange and is proportional to the product $Y_u Y_d$ or $Y_u Y_e$.  Such a diagram is only possible because the SM Higgs doublet is in a self-conjugate representation of the $SU(2)$ gauge group.  In a supersymmetric theory, two Higgs doublets are required since the superpotential containing the Yukawa couplings is holomorphic in the scalar fields as well.\footnote{$H^\dagger H$ can  be written as $\epsilon_{ij} \widetilde H_i H_j$, and so is holomorphic  if $H$ and $\widetilde H$ are considered as independent fields. As a result, whether the Higgs fields occur in holomorphic form is ambiguous.} The Yukawa interaction in Eq.~(\ref{yuk}) no longer contains both $H$ and its conjugate $\widetilde H$,  and the diagrams producing $Y_u Y_{d,e}$ terms do not exist. Note that, for practical purposes, the limit $Y_{u}Y_{e,d}\rightarrow0$ is a good approximation for the SM because the Yukawa matrices are dominated by a single non-zero entry, the top-quark coupling $y_t$.

We can summarize the results so far:  $\dot C_\hol$ respects holomorphy, i.e.\ $\gamma_{\hol \hol}$ is holomorphic ($\gamma_{\hol\overline\hol}=0$) and $\gamma_{\hol \nhol}= 0$, with the exception of two ``non-supersymmetric'' holomorphy-violating terms proportional to $Y_u Y_{d,e}$, which are suppressed phenomenologically. We do not know whether this approximate weak holomorphy is an accident at one loop, or there is something non-trivial going on.

The anomalous dimension matrix also has a large number of vanishing entries in the $\gamma_{\nhol \hol}$ block.  This result suggests a stronger version of holomorphy that seems to be satisfied to a large extent.  In this stronger version, holomorphy in the coefficients $C_\hol$ is imposed on the {\it whole} dimension-six Lagrangian, not only the holomorphic piece.  
This condition requires that the entries in the first column block $\gamma_{i \hol},\ i=\hol,\nhol$ either vanish or are holomorphic. As discussed above, $\gamma_{\hol \hol}$ is holomorphic. The block $\gamma_{\nhol \hol}$ has 24 entries and 21 of them satisfy strong holomorphy. Two of the three entries that violate strong holomorphy (denoted $\strong$) in $\gamma_{\nhol \hol}$ have the $Y_u Y_{d,e}$ form discussed above. Three entries, the contributions to $\psi^2 H^3$ from the $X^2H^2$, $\psi^2 XH$ and $(\overline LR)(\overline LR)$ operators are holomorphic. In this regard $\psi^2 H^3$ operators behave partially as though they should be classified as holomorphic operators. The $H^6-X^2H^2$ entry is the only entry that violates holomorphy when the Yukawa couplings $Y_u Y_{d,e}$ are set to zero.

Since this paper is wildly speculative, we cannot resist the temptation to make one further observation about the lone $\strong$ entry of $\gamma_{\nhol \hol}$ independent of Yukawa couplings. The $H^6-X^2H^2$ non-zero contribution is:
\begin{align}
\dot C_H &= -3 g_2^2 \left(g_1^2+3g_2^2-12 \lambda\right)\mbox{Re} (C_{HW,+})\nn
& -  3 g_1^2 \left(g_1^2+g_2^2-4 \lambda\right) \mbox{Re}(C_{HB,+})\nn
& - 3 g_1g_2 \left(g_1^2+g_2^2-4 \lambda\right)\mbox{Re}( C_{HWB,+})  + \ldots
\label{10}
\end{align}
where the $\ldots$ denote  contributions from non-holomorphic operators.
The $C_{HB,+}$ and $C_{HWB,+}$ terms vanish if $g_1^2+g_2^2=4 \lambda$, i.e.\ if
\begin{align}
m_H^2 = 2m_Z^2 = \left(129\,\hbox{GeV}\right)^2,
\label{m1}
\end{align}
and the $C_{HW,+}$ term vanishes if $g_1^2+3g_2^2=12 \lambda$, i.e.\ if
\begin{align}
m_H^2 = \frac23 m_Z^2  + \frac43 m_W^2 = \left(119\,\hbox{GeV}\right)^2,
\label{m2}
\end{align}
and both terms are highly suppressed near the physical Higgs mass $m_H \sim 126$\, GeV. Since there are two terms of the form Eq.~(\ref{m1}), and one of the form Eq.~(\ref{m2}), the weighted average gives
\begin{align}
m_H^2 &= \frac 23 \left(  2m_Z^2 \right) + \frac 13 \left(  \frac23 m_Z^2  + \frac43 m_W^2\right) \nn
&= 
\frac{14}{9}m_Z^2+\frac49 m_W^2=\left(125.7\,\hbox{GeV}\right)^2,
\label{m3}
\end{align}
which is remarkably close to the measured Higgs mass. At $g_1^2+g_2^2=4 \lambda$, Eq.~(\ref{10}) reduces to
\begin{align}
\dot C_H &= 6 g_1^2 g_2^2\, \mbox{Re} (C_{HW,+})  + \ldots
\label{11}
\end{align}
which has a factor of both the non-Abelian $SU(2)$ and Abelian $U(1)$ gauge couplings. The relation Eq.~(\ref{m1}) is similar to the Higgs mass bound in supersymmetric theories, which arises because the Higgs self-coupling is related to the gauge couplings by supersymmetry.  Note that both Eqs.~(\ref{m1}) and (\ref{m2}) reduce in the custodial $SU(2)$ limit $g_1 \rightarrow 0$ to $g_2^2 = 4 \lambda$.

Finally, the dimension-six operators also contribute to the RGE of the SM parameters~\cite{Jenkins:2013zja,Jenkins:2013wua,Alonso:2013hga}. The gauge couplings run as
\begin{align}
\mu \frac{\rd}{\rd \mu} \left( i\frac{4 \pi}{g_X^2} + \frac{\theta_X}{2\pi} \right) &= \frac{2 m_H^2 }{ \pi g_X^2}i \, C_{HX,+}
\end{align}
where $\theta$-terms are normalized as $\Lag\supset(\theta_Xg_X^2/32\pi^2)X\widetilde X$ and $X \in \left\{SU(3),SU(2),U(1)\right\}$. This equation also respects holomorphy, and the l.h.s.\ is precisely the derivative of the holomorphic gauge coupling in a supersymmetric theory.

\section{Conclusions}
In this paper, we have summarized the non-trivial holomorphic structure of the anomalous dimension matrix for the $d=6$ operators of the SMEFT. Many of the results are similar to those in a supersymmetric theory, even though the SMEFT is not supersymmetric. The weaker form of holomorphy holds up to $Y_u Y_{d,e}$ terms in two entries, and the stronger form of holomorphy is violated in another three entries; two of the three depend on Yukawa couplings  of the form$Y_u Y_{d,e}$. The only non-zero entry which violates holomorphy that is independent of Yukawa couplings leads to Eqs.~(\ref{m1}) and (\ref{m2}), and the Higgs mass value $m_H=125.7$~GeV in Eq.~(\ref{m3}). We have not been able to find a unifying explanation for these results. Given the large number of relations that have to be satisfied for the holomorphic property to hold, it is unlikely to be purely accidental.  There could be a hidden conserved quantum number similar to the conformal spin of $SL(2,R)$ symmetry~\cite{Belitsky:2004cz}, but we have been unable to find one. 

We hope this paper will motivate the community to provide an explanation for the holomorphic structure of the SMEFT RGE at one-loop, and also whether it survives at higher orders. Given the complexity of the SMEFT, it is likely that the results can be extended to more general theories. 

This work was supported in part by DOE grant DE-SC0009919. AM would like to thank A.~Cohen for helpful discussions. We
thank M.B.~Gavela and K.~Intriligator for useful comments on the manuscript.

{\bf Note added:} An explanation for the holomorphic structure has been given by Cheung and Shen~\cite{Cheung:2015aba}. They find that the $\psi^2 H^3-X^2 H^2$ entry should be holomorphic, and we have verified that is indeed the case. We also find that the $\psi^2 H^3-\psi^2 H^3$ entry
is holomorphic up to terms proportional to the Yukawa products $Y_u^\dagger Y_{e,d}^\dagger$. We have updated the table and the discussion in the text to reflect these changes.

\bibliography{hol}

\end{document}